# Memory-built-in quantum teleportation with photonic and atomic qubits

Yu-Ao Chen[1,2], Shuai Chen[1], Zhen-Sheng Yuan[1,2], Bo Zhao[1], Chih-Sung Chuu[1], Jörg Schmiedmayer[3] & Jian-Wei Pan[1,2]

[1]Physikalisches Institut, Universität Heidelberg, Philosophenweg 12, D-69120 Heidelberg, Germany

[2]Hefei National Laboratory for Physical Sciences at Microscale and Department of Modern Physics, University of Science and Technology of China, Hefei, Anhui 230026, People's Republic of China

[3]Atominstitut der Österreichischen Universitäten, TU-Wien, A-1020 Vienna, Austria

**The combination of quantum teleportation[1] and quantum memory[2-5] of photonic qubits is essential for future implementations of large-scale quantum communication[6] and measurement-based quantum computation[7,8]. Even though both of them have been demonstrated separately in many proof-of-principle experiments[9-14], the demonstration of a memory-built-in teleportation of photonic qubits, remains an experimental challenge. Here we demonstrate teleportation between photonic (flying) and atomic (stationary) qubits. In the experiment, an unknown polarization single-photon state is teleported over 7 m onto a remote atomic qubit which serves also as a quantum memory. The teleported state can be stored and successfully read out up to 8 microseconds. Besides being of fundamental interest, our teleportation between photonic and atomic qubits with the direct inclusion of a readable quantum memory**

**represents an important step towards efficient and scalable connection of quantum networks[2-8].**

Quantum teleportation[1], a way to transfer the state of a quantum system from one place to another, was first demonstrated between two independent photonic qubits[9], later developments include demonstration of entanglement swapping[10], open-destination teleportation[11] and teleportation between two ionic qubits[15,16]. Teleportation has also been demonstrated for continuous variable system, i.e. transferring a quantum state from one light beam to another[17] and, most recently, even from light to matter[18].

However, the above demonstrations have several drawbacks, especially in long-distance quantum communication. On the one hand, the absence of quantum storage makes the teleportation of light alone non-scalable. On the other hand, in teleportation of ionic qubits the shared entangled pairs were created locally which limits the distance of teleportation up to a few μm and is difficult to extend to large distances. In continuous variable teleportation between light and matter the experimental fidelity is extremely sensitive to the transmission loss - even in the ideal case only a maximal attenuation of $10^{-1}$ is tolerable[19]. Moreover, the complicated protocol required in retrieving the teleported state in the matter[20] is out of the reach of current technology.

Remarkably, the combination of quantum teleportation and quantum memory of photonic qubits[2-5] could provide a novel way to overcome these drawbacks. Here we achieve this appealing combination by experimentally implementing teleportation between discrete photonic (flying) and atomic (stationary) qubits. In our experiment, we use the polarized photonic qubits as the information carriers and the collective atomic qubits[2-5,12] (an effective qubit consists of two atomic ensembles, each with $10^6$ $^{87}$Rb atoms) as the quantum memory. In memory-built-in teleportation, an unknown polarization state of single photons is teleported onto and stored in a remote atomic qubit via a Bell-state measurement between the photon to be teleported and the photon that is originally entangled with the atomic qubit. The protocol has several distinct features: First, different from ionic system its information carrier (flying photonic qubit) is robust against decoherence and can be easily transmitted over large distances. Second, different from continuous variable system its teleportation fidelity is insensitive to photon losses. In practice, an overall transmission attenuation of $10^{-4}$ is tolerable with current technology, as demonstrated in recent experiments[21,22]. Moreover, since the collective state of atomic ensembles is used to encode an atomic qubit, the teleported state can be easily read out in a controllable time for further quantum information applications. These distinct advantages make our method in principle robust for scalable quantum communication and computation networks[3-5].

A schematic setup of our experiment is shown in Fig. 1. At Bob's site, a pair of effective maximally entangled qubits is created by sending two classical light pulses through two atomic ensembles U (*up*) and D (*down*) which are located in two magneto-optical traps (MOTs) of $^{87}$Rb 0.6 m apart. The two ground states $|a\rangle$ ($5S_{1/2}$, F=2) and $|b\rangle$ ($5S_{1/2}$, F=1) form together with the excited level $|e\rangle$ ($5P_{1/2}$, F'=2) a Λ type system. Initially each ensemble is prepared in the ground state $|a\rangle$. Shining a weak classical write pulse coupling the transition $|a\rangle \to |e\rangle$ with a red detuning Δ (10MHz) and the Rabi frequency $\Omega_W$ into the ensemble *m* (*m=U* or *D*) creates a superposition between the anti-Stokes field $\hat{a}_{AS}$ and a collective spin state of the atoms[2],

$$|\Psi\rangle_m \sim |0_{AS}0_b\rangle_m + \sqrt{\chi_m}|1_{AS}1_b\rangle_m + O(\chi_m), \qquad (1)$$

where $\chi_m \ll 1$ is the excitation probability of one spin flip in ensemble *m*, and $|i_{AS}i_b\rangle$

denotes the *i*-fold excitation of the anti-Stokes field and the collective spin. We adjust $\chi_U = \chi_D$, select orthogonal polarization of the two anti-stokes fields and combine them on a polarized beam splitter (PBS$_1$), as illustrated in Fig. 1. Neglecting the vacuum state and high order excitations, the state of the photonic and atomic qubits can be described as an effectively entangled state

$$|\Psi\rangle = \frac{1}{\sqrt{2}}\left(|H\rangle|\tilde{V}\rangle + |V\rangle|\tilde{H}\rangle\right), \qquad (2)$$

where $|H\rangle/|V\rangle$ denotes horizontal/vertical polarizations of single photons and $|\tilde{H}\rangle = |0_b\rangle_U |1_b\rangle_D$ ($|\tilde{V}\rangle = |1_b\rangle_U |0_b\rangle_D$) denotes one spin excitation in ensemble *D* (*U*). Note that, the anti-Stokes photon has a coherence time of 25 ns[23] and can thus overlap with the photon to be teleported very easily.

After the effectively entangled state (2) is prepared, the anti-Stokes photon is sent to Alice over a 7 m long fibre. Suppose that at Alice's site, the photon to be teleported is in an unknown polarization state $|\phi\rangle = \alpha|H\rangle + \beta|V\rangle$. In terms of four Bell states, $|\Psi^{\pm}\rangle = \frac{1}{\sqrt{2}}\left(|HV\rangle \pm |VH\rangle\right)$ and $|\Phi^{\pm}\rangle = \frac{1}{\sqrt{2}}\left(|HH\rangle \pm |VV\rangle\right)$, the combined state of the three qubits can be rewritten as

$$|\phi\rangle|\Psi\rangle = \frac{1}{2}\left(|\Phi^+\rangle \hat{\sigma}_x |\tilde{\phi}\rangle + |\Phi^-\rangle\left(-i\hat{\sigma}_y |\tilde{\phi}\rangle\right) + |\Psi^+\rangle|\tilde{\phi}\rangle + |\Psi^-\rangle \hat{\sigma}_z |\tilde{\phi}\rangle\right), \qquad (3)$$

where $\hat{\sigma}_x$, $\hat{\sigma}_y$ and $\hat{\sigma}_z$ are the well-known Pauli operators, and $|\tilde{\phi}\rangle = \alpha|\tilde{H}\rangle + \beta|\tilde{V}\rangle$. It can thus be seen that a joint Bell-state measurement (BSM) on the two photons at Alice's side projects the state of atomic qubit at Bob's side into one of the four corresponding states in equation (3). After the BSM, the initial state of photonic qubit is thus transferred to and stored in the atomic qubit. In standard teleportation, depending on the BSM results Bob can then perform a unitary transformation,

independent of $|\phi\rangle$, on the atomic qubit to convert its state into the initial state of the photonic qubit.

To achieve the required BSM, the photon from the entangled state (2) and the photon to be teleported are superposed on a 50:50 beam-splitter (BS in Fig. 1). The BS together with the subsequent coincidence measurements is capable of identifying two of the four Bell-states[24], $|\Psi^+\rangle$ and $|\Psi^-\rangle$ in our experiment. Note that, to demonstrate the working principle of teleportation it is sufficient to identify only one of the four Bell-states, e.g. via identification of $|\Psi^+\rangle$ and verification of $|\tilde{\phi}\rangle$[9-11].

To verify the success of teleportation, we convert the atomic excitation back to optical excitation in a controllable time by shining in two simultaneous read pulses, coupling the transition $|b\rangle \to |e\rangle$ with a blue detuning $\Delta'$ (6MHz) and the Rabi frequency $\Omega_R$. The polarizations of the two read pulses are selected to be perpendicular with respect to the corresponding write pulses. The retrieved Stokes fields are then combined at $PBS_2$. Hence, the atomic qubit is converted back to a single-photon polarization qubit. Instead of performing a direct measurement on the atomic qubit, via a polarization measurement on the converted single-photon state we can thus obtain the experimental teleportation fidelity.

If teleportation occurs, conditional on detecting a $|\Psi^+\rangle$ state at Alice's side the state of the atomic qubit at Bob's side will be left in the state $|\tilde{\phi}\rangle$ (equation (3)). Following the read out protocol the collective atomic state $|\tilde{\phi}\rangle$ will be converted into the initial polarization state $|\phi\rangle$. On the other hand, if a $|\Psi^-\rangle$ state is detected the state of the atomic qubit will then be left in the state $\hat{\sigma}_z|\tilde{\phi}\rangle$, which after conversion is equivalent to the initial state except for a unitary transformation $\hat{\sigma}_z$. Consequently, applying $\hat{\sigma}_z$ on the converted single-photon polarization state we will again obtain the same initial state $|\phi\rangle$. It is worth noting that, the ease of both transferring atomic excitation to optical excitation and exploiting linear optical elements to perform precise unitary transformation on single-photon states is a distinct advantage of our method.

Before performing the teleportation, it is necessary to verify the entanglement. To do so, we map the atomic excitation back into a photon by sending two classical read pulses through the two ensembles. The retrieved Stokes fields with perpendicular polarizations are combined on $PBS_2$ (Fig. 1). And the superposition state of anti-Stokes and Stokes fields is effectively equivalent to the maximally polarization entangled state

$$|\Psi\rangle_{AS,S} \sim |H\rangle_{AS}|V\rangle_S + e^{i(\varphi_1+\varphi_2)}|V\rangle_{AS}|H\rangle_S. \tag{4}$$

Here $\varphi_{1(2)} = \Delta\theta_{W(R)} + \Delta\theta_{AS(S)}$ represents the phase difference between the two anti-Stokes (Stokes) fields at the $PBS_1$ ($PBS_2$). As shown in Fig. 2, the phase shift $\Delta\theta_{W(R)}$ arises from the path difference of the two write (read) beams from $BS_2$ ($BS_1$) to the U and D ensembles; $\Delta\theta_{AS(S)}$ arises from the path difference between the two anti-Stokes (Stokes) fields from the U and D ensembles to the $PBS_1$ ($PBS_2$). In the experiment, $\Delta\theta_W + \Delta\theta_R$ and $\Delta\theta_{AS} + \Delta\theta_S$ are actively stabilized by two Mach-Zehnder interferometers (see Methods), respectively. In this way, $\varphi_1 + \varphi_2$ is actively stabilized and fixed to zero.

With a generation probability of anti-Stokes photon 0.003, the signal-to-noise ratio between the desired ($|H\rangle_{AS}|V\rangle_S$ and $|V\rangle_{AS}|H\rangle_S$) and unwanted ($|H\rangle_{AS}|H\rangle_S$ and

$|V\rangle_{AS}|V\rangle_S$) components is observed to be 15:1, corresponding to a visibility of 87.5% with a statistical error 0.4%. This confirms the $|H\rangle_{AS}|V\rangle_S$ and $|V\rangle_{AS}|H\rangle_S$ terms are two dominant components. Furthermore, in order to prove the two terms are indeed in a coherent superposition, we also measure the polarization correlation in the 45-degree basis. The experimental results exhibit an interference fringe with a visibility of $(82.2\pm0.4)$%, confirming the good quality of atom-photon entanglement.

In the experiment, the initial state to be teleported is prepared using a weak coherent pulse that has the same frequency as anti-Stokes photon. The probability of containing a single photon for each weak coherent pulse is 0.03. Without loss of generality, we select horizontal ($|H\rangle$), 45-degree ($|+\rangle = \frac{1}{\sqrt{2}}|H+V\rangle$) and right-hand circular ($|R\rangle = \frac{1}{\sqrt{2}}|H+iV\rangle$) polarizations as our initial states. As shown in Fig. 1, after knowing the BSM results at Alice's site, the atomic excitation at Bob's site is then converted back to a photonic state in a controllable time to analyze the teleportation fidelity.

With emphasis we note that, since the two-photon events from the weak coherent pulses contribute a significant amount of spurious two-fold BSM coincidences – which herald nothing but the arrival of two source photons and cannot be distinguished from the true BSM results, a two-fold BSM click can only with an average probability of 40% herald the success of teleportation, given an arbitrary initial state. Therefore, as in previous teleportation experiments[9-11], in reality our teleportation only occurs posteriorly, i.e. conditional on detecting a three-fold coincidence. Moreover, due to the imperfect retrieve, collection and detection efficiency of the teleported states, 30%, 75% and 50% respectively (see Methods), in our experiment the overall teleportation success probability is about $10^{-6}$ per experimental run.

Table 1 shows the experimental result of the teleportation fidelities at a retrieve time of 0.5 μs. The result shows the fidelities for different initial states are all well beyond the classical limit of two-thirds, confirming the success of teleportation between photonic and atomic qubits.

To show the ability to store the teleported state in our quantum memory, we further measure the fidelity of teleportation of right-hand circular polarization for different retrieve time. The result is shown in Fig. 3. Up to 8 μs the fidelity is still above the classical limit. The fidelity drops down mainly because of the decoherence in the collective atomic state[25].

In summary, we have demonstrated quantum teleportation between photonic and atomic qubits. The ability - teleporting the unknown quantum state of a photonic qubit onto an atomic qubit and then converting it back to a photonic state at a controllable time - is essential for the recent quantum repeater protocols[3-5] that address the extremely difficult phase stabilization, as required in an original scheme for long-distance quantum communication[2]. However, we would like to note that, due to the low success probability of teleportation and short life-time of quantum memory significant improvements are still needed in order for our method to be useful for practical applications. For example, we could use active feed-forward to achieve both a deterministic entanglement source[3-5] and a high-quality single-photon source[25,26], by which the overall success teleportation rate can be greatly increased while the spurious coincidence is suppressed. This, given our present excitation rate of anti-Stokes photons, would require a lifetime of quantum memory up to 1 ms. Moreover, in order to achieve long-distance quantum communication, e.g. free-space quantum teleportation over 100 km the same order of storage time is required. To do so, one can confine the atoms in an optical trap and exploiting a clock state to store the collective spin excitation[27], this could potentially extend the lifetime up to 1 s. Finally, comparing former photonic teleportation[8-10], where the coherence time of down-converted photons is only about a few hundred fs, the narrowband feature of our anti-Stokes photon source (coherence time ~ 25ns) makes the overlap of independent photon wavepackets from distant sites much easier. This advantage together with the feasible long lifetime quantum memory may provide an ideal solution for large-scale communications, e.g., satellite-based quantum communication[28,29].

**METHODS**

**Experimental cycles.** In the experiment, the MOT is loaded for 20 ms at a repetition rate of 40 Hz. The magnetic field and cooling beams are then quickly switched off while the repumping beams stay on for 0.5 ms before being switched off in order to prepare the atoms in the initial $F = 2$ ground state $|a\rangle$. Then, within another 4.5 ms, experimental trials (each consisting of successive write, read, and repumping pulses) are repeated with a controllable period depending on the desired retrieve time of the teleported state.

**Phase locking.** In order to stabilize the phase $\varphi_1 + \varphi_2$ in expression (4) actively, two Mach-Zehnder interferometers are used as shown in Fig. 2. Because the spatial mode of anti-Stokes (Stokes) field and Write (Read) beam have angle of 3 degrees, we can not lock the phase $\varphi_1 (= \Delta\theta_W + \Delta\theta_{AS})$ and $\varphi_2 (= \Delta\theta_S + \Delta\theta_R)$ directly. However, we can lock the phase of $\Delta\theta_W + \Delta\theta_R$ and $\Delta\theta_{AS} + \Delta\theta_S$ separately. To stabilize the phase of $\Delta\theta_W + \Delta\theta_R$, the read beam is switched on during the 20 ms MOT loading stage, used as the locking beam (Fig. 2a). During the 5ms experimental stage, the shutter is switched off. The interference signal can be used as the error signal of a standard proportional-integrate (PI) locking circuit. The error signal is normalized by the duty cycle and then sent to the homebuilt PI circuit. By controlling the voltage of the piezo (P1) we can lock the phase $\Delta\theta_W + \Delta\theta_R$ to a set value. To stabilize the phase $\Delta\theta_{AS} + \Delta\theta_S$, an additional locking beam polarized at 45 degree with the frequency of read beam is sent in at the angle of the first order diffraction of the AOM (Fig. 2b) during the MOT loading stage. Going through the AOM, the locking beam is overlapped with the Stokes and anti-Stokes beams. Since the anti-Stokes and Stokes light are perpendicularly polarized, the output of the locking beam is from another port of PBS$_1$. After the locking beam goes through a polarizer at 45 degree, the interference signal can be detected by a photodiode and used to lock the phase $\Delta\theta_{AS} + \Delta\theta_S$. During the experimental stage, the shutter and the RF power of AOM are all switched off to prevent the leakage of the locking beam from entering into the anti-Stokes – Stokes channels. In this way, the overall phase of $\varphi_1 + \varphi_2$ is actively locked.

**Noise estimation.** In our experiment, the intensity of the write pulses is adjusted such that in each experimental run the probability of creating an anti-Stokes photon behind the PBS$_1$ is $p_{AS} \sim 0.003$. The intensity of the read pulses is about 70 times higher than the write pulses. Under this condition we achieve a retrieve efficiency of $\gamma \sim 30\%$. After each write and read process, the probability of emitting a single photon in Stokes mode (denoted by $p_S$) is measured to be $\sim 0.004$. In each weak coherent pulse, the probability of containing a single photon is $p_0 \sim 0.03$.

Thus, our 3-fold coincidence would mainly have three components: (1)

Coincidence among a single photon of the initial state from the weak coherent beam, an anti-Stokes photon, and a successfully retrieved Stokes photon, which is the desired event and has a probability of $\sim \frac{1}{2} p_{AS} p_0 \gamma \eta^3$. Here $\eta$ is the average overall detection efficiency of our single-photon detectors, i.e. the product of the collection efficiency ($\sim 75\%$) and the detection efficiency of the detectors ($\sim 50\%$). (2) Spurious coincidence contributed by a two-photon event from the weak coherent pulse and a single-photon event in Stokes mode. In teleportation of $|+\rangle$ and $|R\rangle$ states, the probability of registering such 3-fold coincidence is given by $\frac{1}{4} p_0^2 p_S \eta^3$. However, since only $|\Psi^\pm\rangle$ is analyzed in our BSM, in teleportation of $|H\rangle$ state the two-photon event from the weak coherent beam leads no BSM result and thus no spurious 3-fold coincidence. (3) Spurious coincidence contributed by double emission from the atomic ensembles and one retrieved Stokes photon, which has a probability of $\frac{1}{4} p_{AS}^2 \left(2\gamma\eta - (\gamma\eta)^2\right) \eta^2$.

Moreover, the probability of dark counts in each detector is about $10^{-5}$ per trial, implying a signal-to-noise ratio better than $100:1$. And the errors in polarization are less than 1%. These two errors are thus negligible. Denote the probability of the desired 3-fold coincidence as $S = \frac{1}{2} p_{AS} p_0 \gamma \eta^3$ and the probability of the spurious one as $N = \frac{1}{4} p_0^2 p_S \eta^3 \kappa_\phi + \frac{1}{4} p_{AS}^2 \left(2\gamma\eta - (\gamma\eta)^2\right) \eta^2$ ($\phi$ is the initial state, $\kappa_H = 0$ and $\kappa_+ = \kappa_R = 1$). Taking into account the imperfection of entanglement source, one can thus estimate the final fidelity for $|H\rangle$ teleportation by

$$f = \frac{S(1+V)/2 + N/2}{S+N},$$ where $V \approx 0.88$ is the entanglement visibility in the H/V basis.

A simple calculation shows that the final fidelity should be around 0.90, which is in good agreement with our experimental fidelity $0.865 \pm 0.017$. The slight difference is probably due to the neglected dark count and polarization errors.

In teleportation of $|+\rangle$ and $|R\rangle$ states, the experimental fidelities are much lower. This is because, on the one hand we have one more spurious 3-fold coincidence contribution, i.e. $\frac{1}{4} p_0^2 p_S \eta^3$. More importantly, the imperfect overlap of the wave-packets on the BS, typically around 90% in our experiment[24], will further reduce the fidelities significantly. However, note that such imperfection has no effect on the $|H\rangle$ teleportation fidelity. Taking these into account, a similar calculation shows that the final fidelity for $|+\rangle$ and $|R\rangle$ teleportation should be around 0.79. Given the neglected dark count and polarization errors, our experimental results again well agree with the calculated fidelities.

**Asymmetric systematic error**. Experimentally, the teleportation fidelities of two orthogonal input states under a complete basis could be significantly different due to various asymmetric systematic errors, e.g. caused by asymmetric atom-photon entanglement, state preparation or state analysis. In our experiment, before data collection, various efforts have been made to minimize such asymmetric error. First, the initial polarization state to be teleported is prepared with an extinction ratio better than 500:1. Second, in entanglement preparation and BSM high quality PBSs are selected and adjusted to achieve an extinction ratio better than 300:1 for both H and V polarization. Third, the excitation probability of each ensemble is adjusted properly so that we have a symmetric entangled state (the asymmetric property of the entangled state is verified to be less than 1%). Finally, in polarization analysis of the teleported states an extinction ratio better than 100:1 is achieved by using high quality polarizer. In addition, we have investigated the teleportation fidelities of two states with opposite polarizations in independent experiments and find that the overall asymmetric systematic error of our method is less than 1%.

**Acknowledgements**

This work was supported by the Deutsche Forschungsgemeinschaft (DFG), the Alexander von Humboldt Foundation, the Chinese Academy of Sciences (CAS), the National Fundamental Research Program (Grant No. 2006CB921900) and NNSFC.
.

**Correspondence** and requests for materials should be addressed to Y.-A. C. (yuao@physi.uni-heidelberg.de) or J.-W. P. (jian-wei.pan@physi.uni-heidelberg.de).

**Table 1** Fidelities of teleporting a photonic qubit at a storage time of 0.5 μs. Data for teleporting each state are collected two hours. The error bars represent the statistical error, i.e. ±1 standard deviation.

| Original State | Fidelities |
|---|---|
| $|H\rangle$ | $0.865 \pm 0.017$ |
| $|+\rangle$ | $0.737 \pm 0.009$ |
| $|R\rangle$ | $0.750 \pm 0.009$ |

**Figure Captions:**

**Figure 1 Experimental setup for teleportation between photonic and atomic qubits.** The inset shows the structure and the initial populations of atomic levels for the two ensembles. At Bob's site the anti-Stokes fields emitted from $U$ and $D$ are collected and combined at PBS$_1$, selecting perpendicular polarizations. Then the photon travels 7m through fibres to Alice's to overlap with the initial unknown photon on a beam-splitter (BS) to perform the BSM. The results of the BSM are sent to Bob via a classical channel. Bob then perform the verification of the teleported state in the $U$ and $D$ ensembles by converting the atomic excitation to a photonic state. If a $|\Psi^+\rangle$ is registered, Bob directly performs a polarization analysis on the converted photon to measure the teleportation fidelity. On the other hand, if a $|\Psi^-\rangle$ is detected, the converted photon is sent through a half wave plate (HWP) via the first order diffraction of an AOM (not shown in Figure). The HWP is set at 0 degree serving as the unitary transformation of $\hat{\sigma}_z$. Then the photon is further sent through the polarization analyzer to obtain the teleportation fidelity.

**Figure 2 Schematic drawing of the phase locking setup.** Two Mach-Zehnder

interferometers are used to actively stabilize the phases between the arms of write and read paths (**a**) and between the arms of anti-Stokes and Stokes paths (**b**), respectively. H/V denotes the horizontal/vertical polarization, and AOM is for an acousto-optic modulator. A polarizer (Pol.) is set at $45^0$ to erase the polarization information. The $\lambda/2$ plates are set at $45^0$ as well to rotate the horizontal polarization to vertical. AS (S) denotes the anti-Stokes (Stokes) photon.

**Figure 3 Fidelity of $|R\rangle$ teleportation along storage time.** Until 8 μs the fidelity is still well beyond the classical limit of 2/3. Each experimental point is measured for about four hours (averagely). The curve is a Gaussian fit, due to the Gaussian decay of the retrieve efficiency. The error bars represent the statistical error, i.e. ±1 standard deviation.

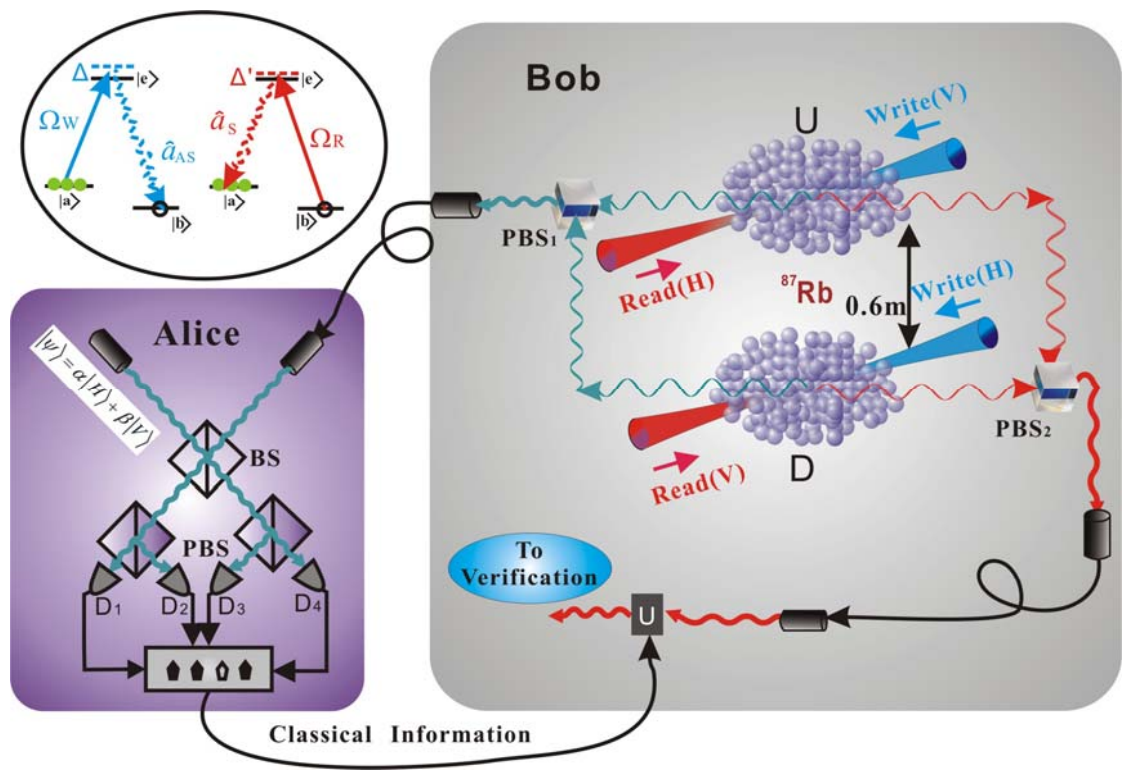

**Figure 1**

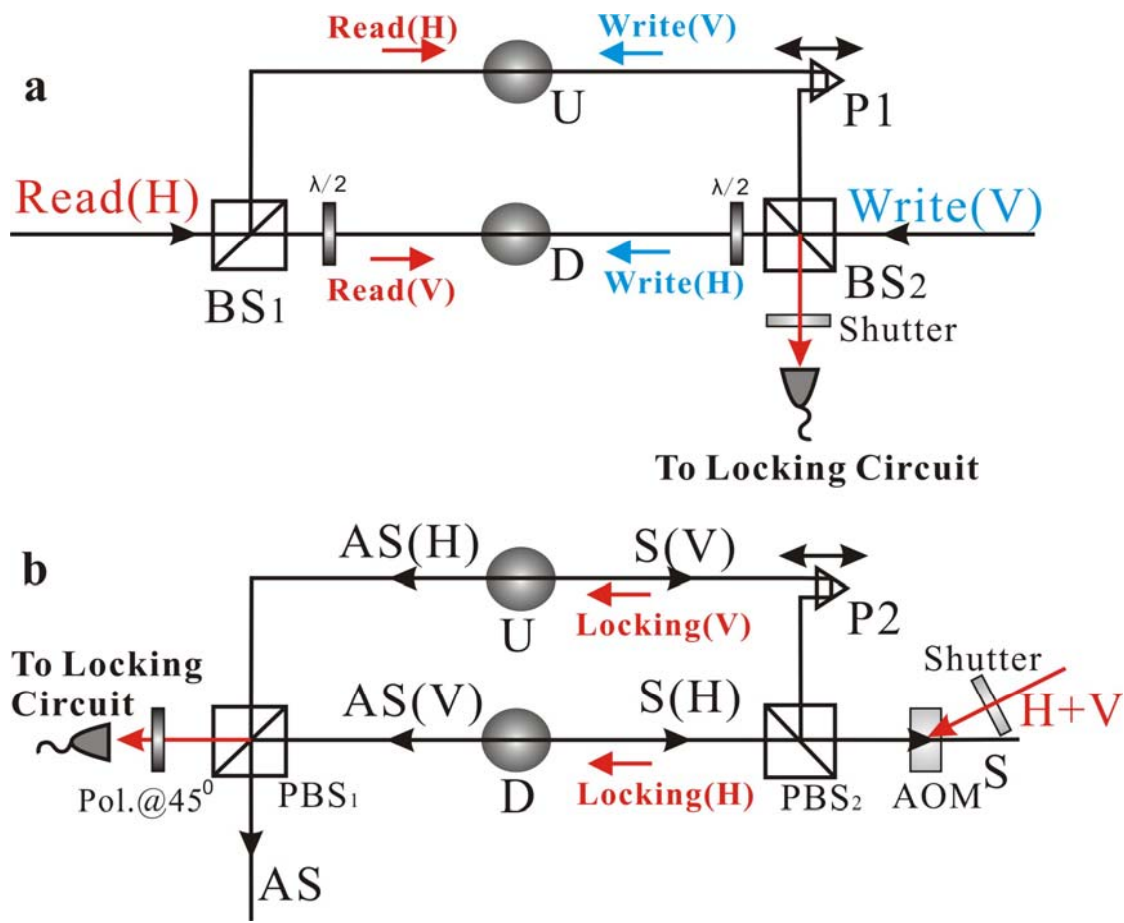

**Figure 2**

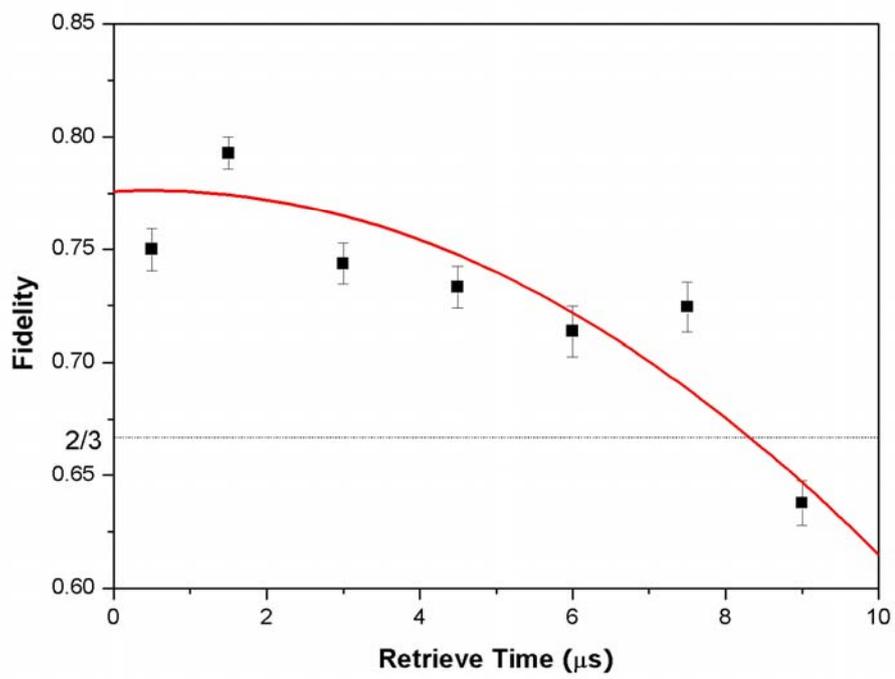

**Figure 3**